# Ozone Layer and the Formation of the Ozone Hole

A.A. Mikhailov

Yu.G. Shafer Institute of Cosmophysical Research and
Aeronomy, 31 Lenin Ave., 677891 Yakutsk, Russia

> A hole frightens. The hole is one
> of the meaning which we still will
> have to understand. And also it
> means that when you see a bullet,
> flying to you,then it is already
> late to run away.    K.Moore.

**Abstract**

The current state of this problem has briefly been considered. The
hypothesis on the formation of the ozone hole due to the action
of some molecules emitted by the rockets has been suggested.

## 1.  Introduction

A way of contaminating substances from the troposphere into the stratosphere is by no means simple. The laws of atmospheric dynamics are that the transport of the air mass across the tropopause does practically not occur. Therefore, the small component molecules cannot penetrate from the troposphere into the stratosphere moving directly upwards. The very cold and large-altitude (17-18 km) tropopause of tropical latitudes, where that transport is takes place, is the one exception.  As a result, the small component molecules, penetrated in the troposphere of the middle latitudes, must pass the long way: at first in the troposphere to the equator (horizontal transport), then through the tropical tropopause(vertical transport) and, at last, backwards, to middle latitudes but at stratospheric altitudes(horizontal transport). Note, in the way to the equator the greater part of small component molecules is returned back to the ground together with along the precipitation. That is why no significant decrease of the ozone layer from volcanoes and earthquake last centuries was observed.

Apparently, the addition to the above mechanism is the penetration of powerful cumulus clouds into the atmosphere, which are usually formed in the Antarctic. In this case, the movement of air inside of clouds transfers the small admixtures immediately into the stratosphere.

## 2. Observations

First information on the detection of a deficit of total amount of ozone above the Antarctic dates to 1979. The considerable decrease of ozone is observed every year in October, i.e. in the period of the Antarctic spring. The total amount of ozone decreases up to 50% in different places [2]. In August 1987 the record depleted amount of ozone was registered, the ozone layer thickness was only 100 u.D. [2] .

A region of the considerable ozone depletion is named an ozone hole. Recently the depleted ozone concentration was detected above Arctic, Western Europe, South America and Eastern Siberia. For example, the ozone layer thickness above the Great Britain, at Lerwick station (Scotland Islands) on March 5, 1996 was 195 u.D.[2] .

A correlation between the ozone hole and one of small component, in particular, the chlorine oxide has been established. In the ozone hole region above the Antarctic the chlorine oxide was more by a factor of 120 than beyond the ozone hole.

Decribe briefly a popular hypothesis to form the ozone holes above the South Pole in spring. In winter in the polar stratosphere of the Southern Hemisphere a stable cyclone exists, what is known as the circumpolar vortex. In air inside the vortex moves mainly along closed trajectories not going out of its boundaries. By the late winter it cools strongly (up to -80°C) and the polar clouds appear in the stratosphere, which consist of the ice crystals and drops of supercooling liquid. The particles of polar clouds connect the nitrogenous compounds and give scope to the action of chlorine cycle to destruct ozone. As the Antarctic stratosphere is warming up, the circumpolar vortex is destructed. In this case, an exchange with air richer by ozone at middle latitudes is restored, the stratospheric clouds disappear, released molecules of nitrogenous compounds connect the chlorine oxide molecules, distracting action of the chlorine cycle to ozone become weaken, and the ozone amount is restored to its unperturbed values.

If chlorine penetrates somehow into the stratosphere, then under its interaction with ozone the chlorine oxide is produced, and further a chain of destruction of ozone is occurred. One molecule of the oxide may destruct up to 100 thousand molecules of ozone.

The ground measurements show that the total amount of ozone for 1969-1986 is reduced, on the average, by 2-3% [2] . The ultraviolet (UV-B) of wavelength $\lambda = 250$ nm is decreased due to its absorption in the ozone layer by a factor of $10^{17}$(one with 17 zeros) when the thickness of average ozone layer (300 u.D.) is decreased to 250 u.D., the ultraviolet intensity is increased by a factor of 600 (the major danger is that the ultraviolet intensity is increased as an exponential law, see Figure), to 200 u.D. - by 400 000 times; to 150 u.D. - by 300 billion times; to 100 u.D. - by 200 billion times. We for the present not realize what danger is from the depletion of the ozone layer.

As it was reported in newspapers several years ago, one fine day above the southern region of Chile, near Punta-Arenas the ozone hole is appeared. As a result of the ultraviolet action, the hundreds of cows and sheep's became blinded, people caught burns. Also, the decrease of the total thickness of the ozone layer will cause the noticeable warming of climate, because ozone takes an active part in the creation of a frame effect. Thereby, the significant changes of weather in some regions of the Earth, floods, a higher of the ocean water level and other very serious consequences would be expected. As a result of people activities, more and more pollution substances penetrate into the atmosphere. A part of these substances is harmful for ozone, promoting its destruction. Among these are the compounds of nitrogen, hydrogen and chlorine. Nitrogen penetrates into the atmosphere in the shape of nitrogenous oxides, when nitrogenous fertilizers are used and at an ejection of spent gases in flights of high-altitude airplanes. The enterprises emitting methane into the atmosphere and also the flights of the high-altitude airplanes and rockets are the major contributes of hydrogen compounds. Chlorine penetrates into the atmosphere as a result of the use of chlorine's organic components, first of all freon, in the home and industrially.

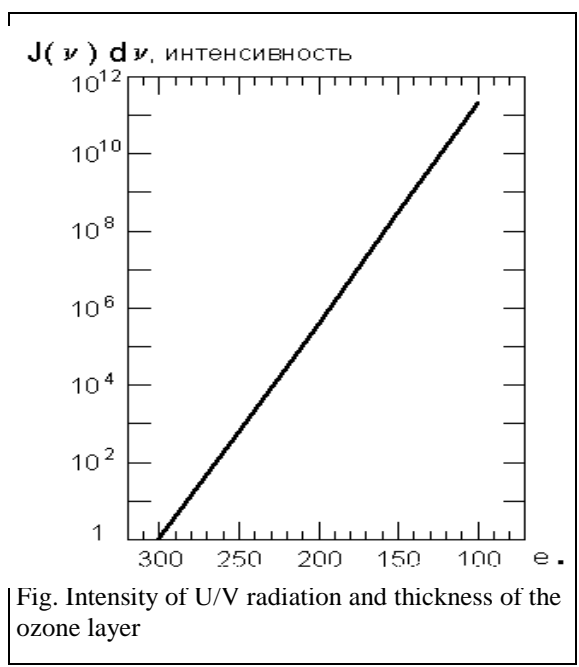

Fig. Intensity of U/V radiation and thickness of the ozone layer

Freon is a chlorine-fluorine-carbons compound. It is used in the refrigerators, in the manufacture of aerosols (deodorants, varnishes, insecticides etc.) and also in different fields of technology (greases, the preparation of foam etc.).

The scientists suppose that chlorine compounds are of the greatest danger for ozone, in the stratosphere. The ejections of chlorine's substances are catastrophically increased in the last decades. For example, the ejection of freon -11 increased by a factor of 300 for 1950-1980 [2]. The lifetime of it, up to the disappearance by the solar emission action or chemical reaction, is 70-100 years, on the average. As noted above, contaminating substances emitted into the atmosphere at middle latitudes, where the preponderance of enterprises, do not penetrate at once into the stratosphere. The calculations show that substances penetrating into the atmosphere to date, will penetrate into the stratosphere, on the average, approximately in 100 years, but freons by now are detected in the stratosphere. The manner in which they penetrated so quickly into the stratosphere, if their preparation has begun only 1930's, and mass preparation - in the middle of 1960's remains a mystery. Some scientists assume [2] that freons are no major reason to form the ozone holes(if this is not the case, they are no more than 30%).

### 3. Rockets

In my opinion, the rockets are the major reason that dangerous for ozone substances penetrate into the stratosphere, and a source of the ozone layer depletion. A shock wave starting from a rocket or falling finished off stage can form a peculiar corridor between the troposphere and stratosphere. For example, when the American rocket "Atlas" was launched, a "hole" (the electron flux was as small as one order of magnitude in a circle of diameter of several hundreds of kilometers) was registered in the region of ionosphere F-layer. Along that corridor the dangerous for ozone substances may directly penetrate into the stratosphere omitting a process of return back on the ground together with the precipitation on their way to equator. A perchlorate, which is a source of chlorine and a part of solid fuel, penetrates into the stratosphere directly from rockets.

Water of great quantity and other hydrogen oxides are also ejected from the rockets. Unfortunately, I not know exactly what other dangerous for ozone substances are ejected from rockets. Note the rocket is usually of the oblique trajectory, which extending in the atmosphere several thousands of kilometers at altitudes of 100-200 km, ejects the gas sprays above the stratosphere.

While the rocket motor is running, the velocity of flux of combustion product from a nozzle is 3-4 km/s and temperature is 3000°C. A mass of combustion products is up to 300 kg/s. When it is considered, that the rocket motors operate more than 1 hour, then in each launching of the rocket more than 1000 tons of combustion products are ejected into the atmosphere. Under the normal regime of operating motor, combustion products are of the gaseous phase. But in the period of the start or stop of the motor (transitional operating regime) the combustible is not consumed completely, and aerosols are beginning to form. At the separation of the finished off rocket stage the combustible remainders, what is known as a guaranteed reserve (up to 1-2 % of refueled mass), are poured out from the tanks through drain-pipes into the atmosphere. The ejection is also of the third stage and the braking of the rocket by means of the decrease of pressure in the combustion camera. Those dangerous for ozone substances propagate by means of the shock wave, high- altitude wind and latitudinal drift to vast distances up to the Earth's poles. Arguments in favor of the above hypothesis are:

1. Investigations of the ozone layer above the Antarctic shown that the significant decrease of the ozone concentration is observed in the outer stratosphere. This demonstrates that the dangerous for ozone substances penetrate into the stratosphere from above. The only rockets pollute the air at the altitudes of 30-200 km.

2. It is found that in the period between 1969 and 1993 in the stratosphere between latitudes of 45° N and 65° N the ozone concentration was decreased by about 14% [2]. That latitude range coincides with a flight corridor of rockets launched from the Northern Hemisphere.

3. The large size ozone holes were detected over Yakutia and South America where the production and mass application of freon absent. But namely above those regions of the Earth over prolonged periods the second and third stages of Russian rockets fallen.

4. When the American solid-fuel rocket "Space Shuttle" is launched, up to 10000 tons of ozone are destructed [3]. In the Earth's atmosphere there are only 3 billion tons of ozone. The fact, that the launching of rockets is growing year after year, is disturbing. In recent years, most of launchings are realized with the commercial goal.

Thus, the American Company "Iridium" realizes already a project of the installation for telecommunications of 69 satellites (to profit by international talks). Other project "Teledesik-Internet on Sky" is developed for which 40 satellites are required. In 1995 20 commercial satellites were launched all over the world, in 1997 it is planed to launch 76 satellites, in 1998-121. If we are not want to lost, it is necessary to decrease by this time the number of launchings of rockets, making them only as a last resort. The rocket launchings are to be stopped above high latitudes where the greatest ozone concentration is observed, and in the future all launching must be made from the equator along the vertical trajectories. It is desirable to launch the monolithic rockets without the stages after the ecological expert opinion.

## 4. Conclusions

If we now take no precautions, the following situation can occur. After a time, extraterrestrial people will arrive on our lifeless Earth and see a placard "No rockets!" One of the new-comers will say the another "Evidently, they realized the danger starting from the rockets, but here just one of the ozone holes covered them".